\documentclass[prl,a4paper,10pt,twocolumn,superscriptaddress,amsmath,amsfonts,amssymb,preprintnumbers,citeautoscript]{revtex4}
\pdfoutput=1
\usepackage[T1]{fontenc}\usepackage[latin1]{inputenc}
\usepackage{graphicx,color,booktabs,microtype,afterpage,sidecap}
\usepackage[autoplay,palindrome]{animate}
\usepackage[charter]{mathdesign}

\renewcommand{\sidecaptionsep}{2em}
\usepackage[colorlinks,plainpages=false,linkcolor=black,urlcolor=blue,citecolor=black,pdfpagemode=UseNone,pdfstartview=FitBH]{hyperref}

\makeatletter
\renewcommand{\@evenfoot}{\hfill\bf\thepage}
\renewcommand{\@oddfoot}{\hfill\bf\thepage}
\makeatother

\makeatletter
\renewcommand{\@evenfoot}{\hfill\bf\thepage}
\renewcommand{\@oddfoot}{\hfill\bf\thepage}
\makeatother

\newcommand{ \ybco }{\mbox{YBa$_2$Cu$_3$O$_{6+x}$}}

\newcommand{ \ybcosixsix }{\mbox{YBa$_2$Cu$_3$O$_{6.6}$}}

\newcommand{ \qq }{\mbox{$\mathbf{Q}$}}
\newcommand{ \qaf }{\mbox{$\mathbf{Q}_\textup{AFM}$}}

\newcommand{ \Tc}{\mbox{$T_{\text{c}}$}}
\newcommand{ \deltaSC}{\mbox{$\Delta$}}
\newcommand{ \Eres}{\mbox{$\hslash\omega_\textup{res}$}}
\newcommand{ \astar}{\mbox{$a^*$}}
\newcommand{ \bstar}{\mbox{$b^*$}}
\newcommand{ \cstar}{\mbox{$c^*$}}

\begin{document}


\title{Normal-State Spin Dynamics and Temperature-Dependent Spin Resonance Energy\\in an Optimally Doped Iron Arsenide Superconductor}

\author{D.\,S.\,Inosov}
\affiliation{Max-Planck-Institut für Festkörperforschung, Heisenbergstraße 1, 70569 Stuttgart, Germany}

\author{J.\,T.~Park}
\affiliation{Max-Planck-Institut für Festkörperforschung, Heisenbergstraße 1, 70569 Stuttgart, Germany}

\author{P.~Bourges}
\affiliation{Laboratoire L\'{e}on Brillouin, CEA-CNRS, CEA Saclay, 91191 Gif-sur-Yvette Cedex, France}

\author{D.\,L.~Sun}
\affiliation{Max-Planck-Institut für Festkörperforschung, Heisenbergstraße 1, 70569 Stuttgart, Germany}

\author{Y.~Sidis}
\affiliation{Laboratoire L\'{e}on Brillouin, CEA-CNRS, CEA Saclay, 91191 Gif-sur-Yvette Cedex, France}

\author{A.~Schneidewind}
\affiliation{Institut für Festkörperphysik, Technische Universität Dresden, D-01062 Dresden, Germany}
\affiliation{Forschungsneutronenquelle Heinz Maier-Leibnitz (FRM-II), TU München, D-85747 Garching, Germany}

\author{K.~Hradil}
\affiliation{Institut für Physikalische Chemie, Universität Göttingen, 37077 Göttingen, Germany}
\affiliation{Forschungsneutronenquelle Heinz Maier-Leibnitz (FRM-II), TU München, D-85747 Garching, Germany}

\author{D.\,Haug}
\affiliation{Max-Planck-Institut für Festkörperforschung, Heisenbergstraße 1, 70569 Stuttgart, Germany}

\author{C.\,T.~Lin}
\affiliation{Max-Planck-Institut für Festkörperforschung, Heisenbergstraße 1, 70569 Stuttgart, Germany}

\author{B.~Keimer}
\affiliation{Max-Planck-Institut für Festkörperforschung, Heisenbergstraße 1, 70569 Stuttgart, Germany}

\author{V.~Hinkov}
\affiliation{Max-Planck-Institut für Festkörperforschung, Heisenbergstraße 1, 70569 Stuttgart, Germany}

\begin{abstract}
\center\bigskip\thispagestyle{plain}
\begin{minipage}{\textwidth}\textbf{The proximity of superconductivity and antiferromagnetism in the phase diagram of iron arsenides \cite{ChuAnalytis09, DrewNiedermayer09}, the apparently weak electron-phonon coupling \cite{BoeriDolgov08} and the ``resonance peak'' in the superconducting spin excitation spectrum \cite{ChristiansonGoremychkin08, ChiSchneidewind09, LumsdenChristianson09, LiChen09} have fostered the hypothesis of magnetically mediated Cooper pairing. However, since most theories of superconductivity are based on a pairing boson of sufficient spectral weight in the normal state, detailed knowledge of the spin excitation spectrum above the superconducting transition temperature \Tc\ is required to assess the viability of this hypothesis \cite{Parks69, DahmHinkov09}. Using inelastic neutron scattering we have studied the spin excitations in optimally doped BaFe$_{1.85}$Co$_{0.15}$As$_2$ ($T_\text{c}=25$\,K) over a wide range of temperatures and energies. We present the results in absolute units and find that the normal state spectrum carries a weight comparable to underdoped cuprates \cite{FongBourges00, LipscombeVignolle09}. In contrast to cuprates, however, the spectrum agrees well with predictions of the theory of nearly antiferromagnetic metals \cite{Moriya85}, without complications arising from a pseudogap \cite{NormanPines05, RossatMignod91, LeeYamada03} or competing incommensurate spin-modulated phases \cite{TranquadaSternlieb95}. We also show that the temperature evolution of the resonance energy follows the superconducting energy gap \deltaSC, as expected from conventional Fermi-liquid approaches \cite{KorshunovEremin08, MaierGraser09}. Our observations point to a surprisingly simple theoretical description of the spin dynamics in the iron arsenides and provide a solid foundation for models of magnetically mediated superconductivity.}\end{minipage}
\end{abstract}

\maketitle\thispagestyle{empty}\clearpage

In conventional superconductors such as mercury or niobium, the electron system gains energy by establishing a superconducting condensate consisting of Cooper pairs bound by the exchange of virtual phonons. Other elementary excitations also have the potential to mediate pairing: In heavy-fermion superconductors like CeCoIn$_5$ or UPd$_2$Al$_3$, antiferromagnetic (AFM) spin excitations are likely to be involved in the pairing mechanism \cite{SatoAso01}. However, coupling of itinerant carriers to the quasi-localized rare-earth $f$\!-electrons introduces a complexity that has hitherto precluded a commonly accepted theory \cite{SatoAso01}.

In cuprate high-$T_\text{c}$ superconductors, AFM spin excitations are also among the most promising contenders for the pairing boson \cite{Eschrig06}, despite the remaining controversy about the role of electron-phonon interactions. Here, the complication comes from strong electron interactions in the form of on-site Coulomb repulsion, which render the parent compounds AFM Mott insulators, and from a multitude of poorly understood phenomena, such as the normal-state pseudogap and the competition of superconductivity with incommensurate spin- and charge-modulated phases \cite{TranquadaSternlieb95}. Even the adequacy of boson-mediated pairing schemes itself has been called into question \cite{Anderson07}.

\begin{figure*}[t]
\includegraphics[width=0.85\textwidth]{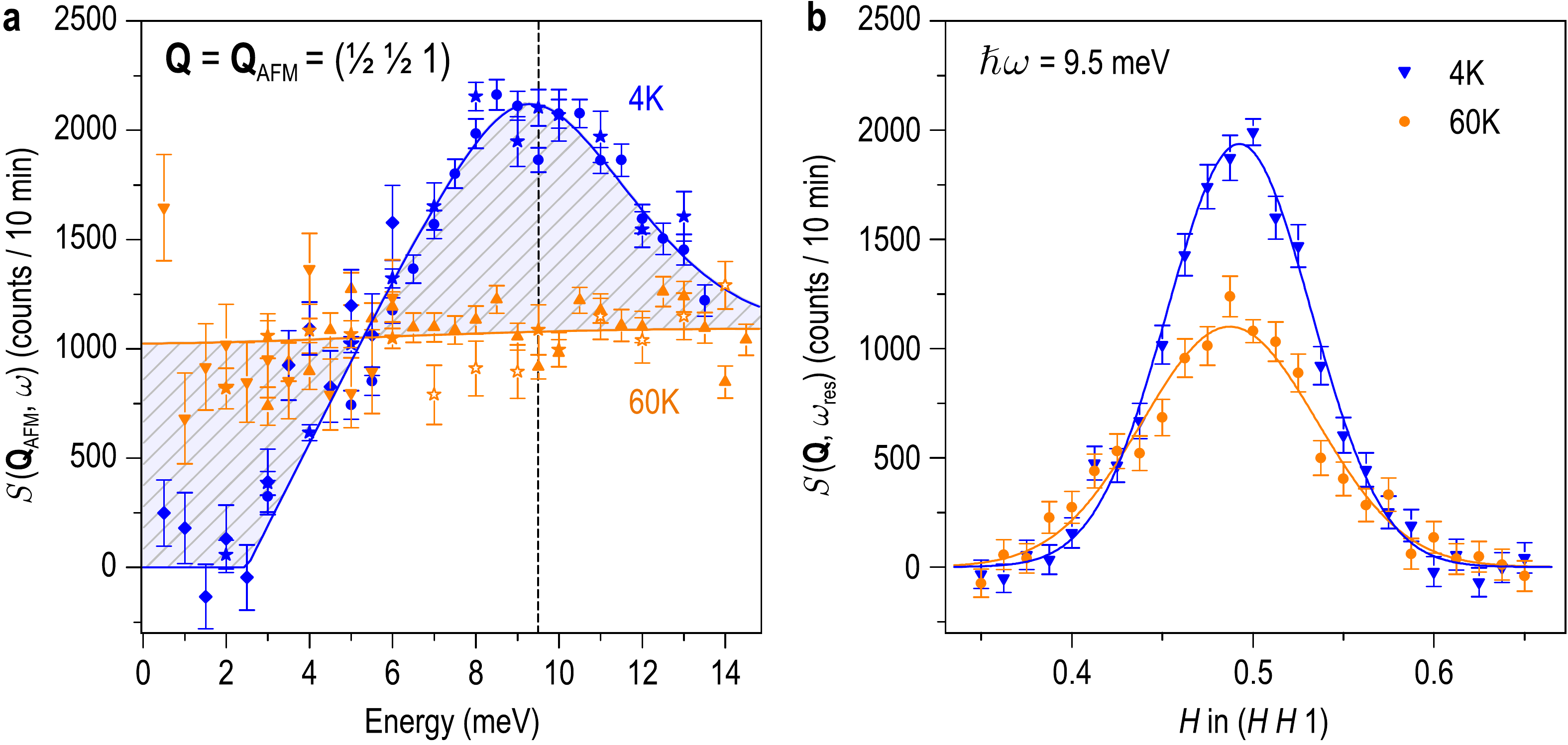}
\caption{Spin excitations in the vicinity of the AFM wavevector \qaf, in the superconducting ($T=4$\,K) and the normal state ($T=60$\,K). \textbf{a},~Energy evolution of the magnetic scattering function $S(\qaf,\omega)$ after a background correction (see Methods). Different symbol shapes represent measurements at different spectrometers. Solid lines are guides to the eye. \textbf{b},~Wavevector dependence of $S(\qq,\omega)$ measured at the resonance energy (dashed line in panel \textbf{a}). A linear background has been subtracted. The lines are Gaussian fits.}
\label{RawEdep}
\end{figure*}

The recently discovered iron arsenide superconductors \cite{KamiharaWatanabe08, SefatJin08} are characterized by AFM correlations throughout the phase diagram, often coexisting with superconductivity deep into the superconducting dome \cite{DrewNiedermayer09}. Besides, it was shown \cite{BoeriDolgov08} that electron-phonon coupling is too weak to explain the high $\Tc$, which turns the spotlight onto the magnetic coupling channel again \cite{MazinSingh08, KurokiOnari08}. While iron arsenides also derive from AFM parents, unlike cuprates they remain metallic at all doping levels, rendering Fermi-liquid based approaches more promising than in cuprates.

In several of these unconventional superconductors, a redistribution of AFM spectral weight into a ``resonance peak'' at an energy $\hslash\omega=\Eres$ smaller than the superconducting gap $2\Delta$ heralds the onset of superconductivity \cite{SatoAso01, RossatMignod91, FongKeimer96}. Since the intensity of this mode is determined by coherence factors in the superconducting gap equation, it is only expected to occur for particular gap symmetries and was one of the first indications for $d$-wave superconductivity in the cuprates. The recent discovery of a resonant mode in both hole-doped Ba$_{1-x}$K$_x$Fe$_2$As$_2$ \cite{ChristiansonGoremychkin08} and electron-doped BaFe$_{2-x}$(Ni,\,Co)$_x$As$_2$ \cite{ChiSchneidewind09, LumsdenChristianson09,LiChen09} is therefore an important achievement. While the existence of a resonance was shown to be compatible with a sign-reversed $s_\pm$-wave superconducting gap \cite{KorshunovEremin08, MaierGraser09}, it is a generic consequence of the opening of the gap and hence does not \emph{per se} constitute evidence of a magnetic pairing mechanism.  Since a pairing boson of sufficient spectral weight must be present already above \Tc, detailed knowledge of both the spectrum in the normal state and its redistribution below $T_\text{c}$ is a prerequisite for a quantitative assessment of theoretical models, as recently demonstrated for \ybcosixsix\ \cite{DahmHinkov09}.

Here we study the spin excitations in a single crystal of optimally electron-doped BaFe$_{1.85}$Co$_{0.15}$As$_2$ (\Tc\,=\,25\,K) at temperatures up to $T=280$\,K and energies up to $\hslash\omega=32$\,meV (>\,$4\deltaSC$). We begin by showing in Fig.\,1a the scattering function $S(\qq,\omega)$ at the antiferromagnetic wavevector $\qq=\qaf=\bigl(\frac{1}{2}\,\frac{1}{2}\,1\bigr)$ for $\hslash\omega\leq15$\,meV in the superconducting state (4\,K) and in the normal state (60\,K). The data were obtained by collecting a series of \qq-scans at fixed $\omega$, and $\omega$-scans at fixed \qaf, supplemented by points appropriately offset from $\qaf$ to allow an accurate background subtraction. We determine \Eres\ to be 9.5\,meV, in agreement with previous investigations on samples of similar doping levels \cite{LumsdenChristianson09}. At this stage, we present $S(\qq,\omega)$ instead of the dynamical susceptibility $\chi''(\qq,\omega)$, since a sum rule holds, stipulating that $\int_{-\infty}^{\infty}\mathrm{d}\omega\int\!\mathrm{d}\qq\,S(\qq,\omega)$ is $T$-independent. An important result is that within the experimental error the resonant spectral-weight gain is compensated by a depletion at low energies, and that the superconductivity-induced effects are limited to $\hslash\omega$\,$\lesssim$\,$2\deltaSC$ (see also Fig.\,2). The \qq-integration can be neglected here, since within the shown energy range of up to $2\Delta$ the spectrum remains commensurate and the measured \qq-width does not change appreciably (Fig.\,1b and Supplementary Information): Its value of $\sim$\,0.1\,r.l.u. is much broader than the resolution and thus represents the intrinsic \qq-width to a good approximation.

We next obtain $\chi''(\qaf,\omega)$ by correcting $S(\qaf,\omega)$ for the thermal population factor, which is largest at low $\omega$ and high $T$ (Fig.\,2). Performing this correction, we now clearly establish that the low-$\omega$ depletion represents a real spin gap (not to be be confused with the superconducting gap $\deltaSC$) and not a trivial thermal population effect. One of the central results of our study is that we can present $\chi''(\qq,\omega)$ in absolute units (see Methods). Apart from its importance for theoretical work, this allows us to extract the weight of the spectral features to be discussed in the following.

\begin{SCfigure*}[][t]
\includegraphics[width=0.6\textwidth]{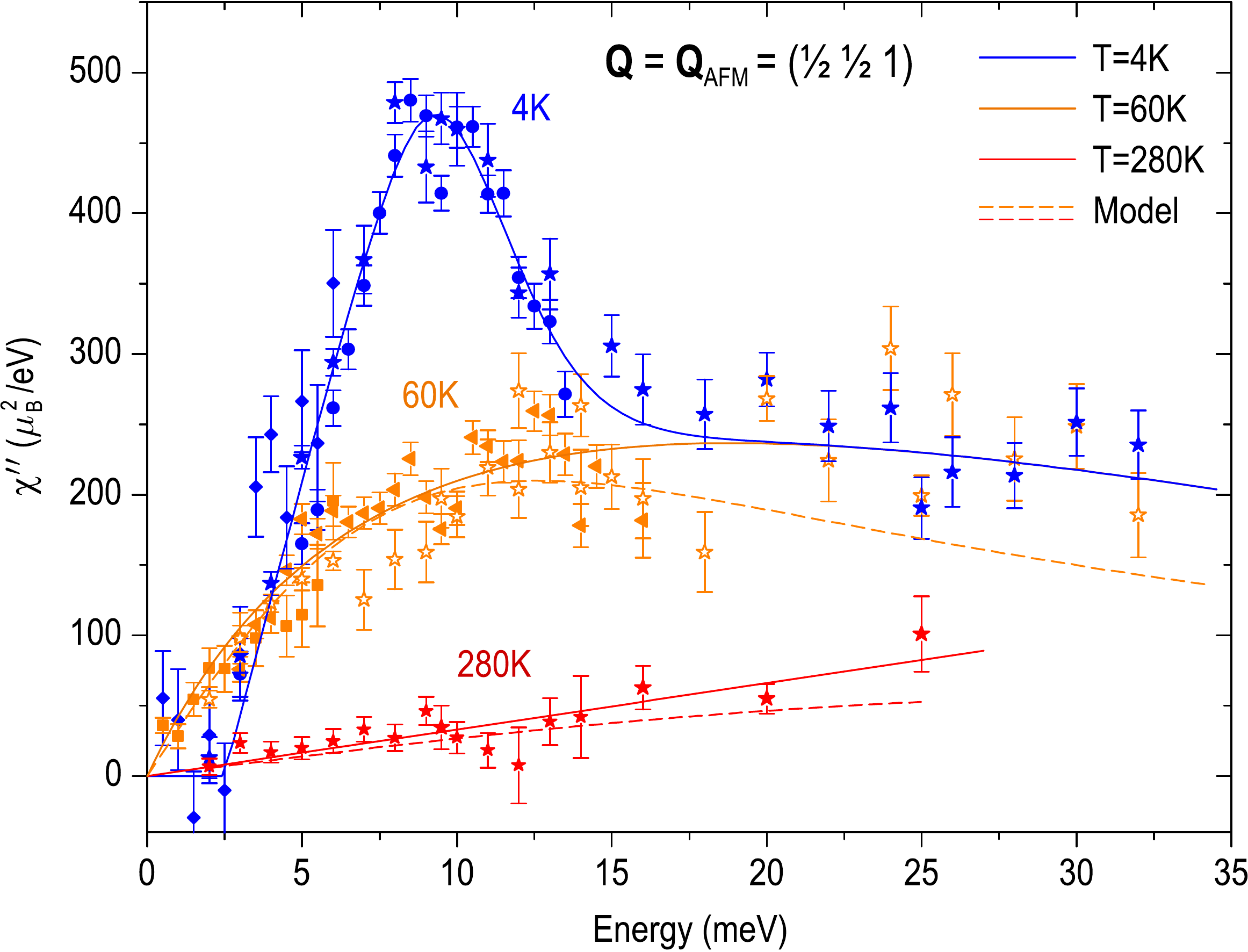}
\caption{Imaginary part of the spin susceptibility $\chi(\qaf,\omega)$ in the superconducting ($T=4$\,K) and the normal state ($T=60$ and 280\,K). The data were obtained from $S(\qq,\omega)$ by correcting for the thermal population factor and were put on an absolute scale as described in the Supplementary Information. The solid lines are guides to the eye. The dashed lines represent global fits of the formula described in the text to all the normal state data in this figure and Fig.\,3.\vspace{15em}}
\label{Edep}
\end{SCfigure*}

\begin{figure*}[t]
\includegraphics[width=0.9\textwidth]{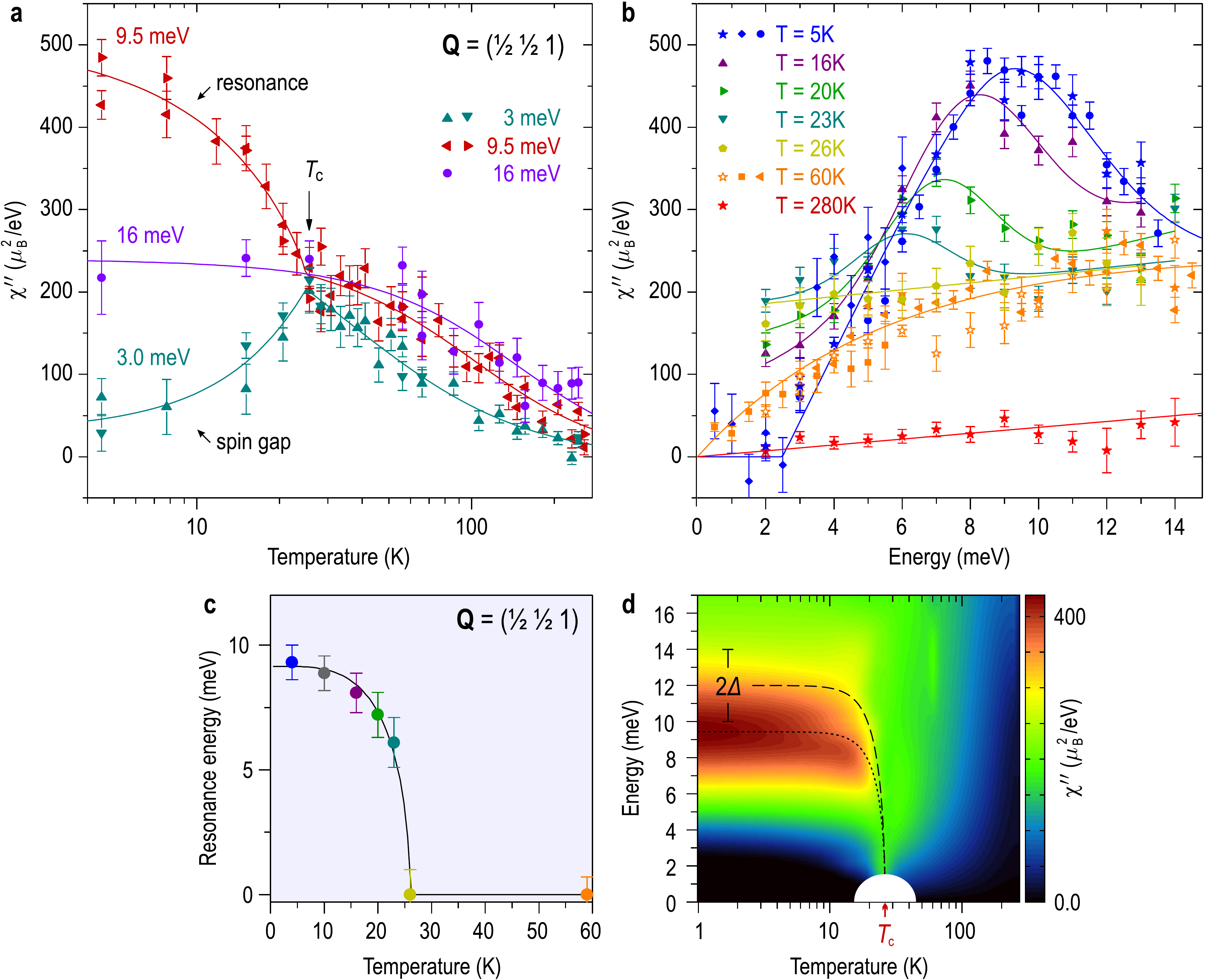}
\caption{Energy and temperature dependence of $\chi''(\qaf,\omega)$ and evolution of the resonance peak below \Tc. \textbf{a},~Temperature dependence of $\chi''(\qaf,\omega)$ at three different energies: within the spin gap (3\,meV), at \Eres\ (9.5\,meV) and above $2\deltaSC$ (16\,meV). \textbf{b}, Energy scans at \qaf\ showing $\chi''(\qq,\omega)$ at different temperatures. The lines in \textbf{a} and \textbf{b} are guides to the eye. \textbf{c}, Temperature evolution of the resonance energy $\Eres(T)$ defined by the maxima in panel \textbf{b}. The line has the same functional dependence as the superconducting gap \deltaSC\ obtained by angle-resolved photoemission \cite{EvtushinskyInosov09, TerashimaSekiba09}, that is $\omega_\text{res}(T)\propto\deltaSC(T)$. \textbf{d}, Interpolation of the data in panels \textbf{a} and \textbf{b} showing $\chi''(\qaf,\omega)$ in the $\omega$-$T$ plane for $T$ up to 280\,K. The vertical bar shows the interval of the reported $2\deltaSC$ values \cite{SamuelyPribulova09, YinZech09, TerashimaSekiba09}. The dotted line is the same as the fit in \textbf{c}. The dashed line has the same functional dependence and tracks the average value of $2\deltaSC(T)$ as a function of $T$. Note the logarithmic $T$-scale in panels \textbf{a} and \textbf{d}.\vspace{5pt}}
\label{Tdep}
\end{figure*}

In the normal state at 60\,K we observe a broad spectrum of gapless excitations with a maximum around 20\,meV and a linear $\omega$-dependence for $\omega\rightarrow0$. Increasing $T$ to 280\,K suppresses the intensity and presumably shifts the maximum to higher energies, while the low-energy linearity is preserved. This behavior and the absence of complications by incommensurate modulations or a pseudogap (see also Fig.\,3a) motivates an analysis within the framework of the theory of nearly antiferromagnetic Fermi liquids \cite{Moriya85}, for which
\begin{equation}
\chi''_{\kern-.5pt T}(\mathbf{Q},\,\omega)=\frac{\chi_{\kern-.5pt T}\kern.5pt\Gamma_{\kern-.5pt T}\,\omega}{\omega^2\kern.5pt+\Gamma_{\kern-.5pt T}^{\kern.5pt2}\kern.5pt\bigl(1+\xi_{\kern-.5pt T}^2|\mathbf{Q}-\mathbf{Q}_\text{AFM}|^2\bigr)^2}.
\label{Eq:Moriya}
\end{equation}
Here $\chi_{\kern-.5pt T}=\chi_0\,(T+\Theta)^{-1}$ represents the strength of the AFM correlations in the normal state, $\Gamma_{\kern-.5pt T}=\Gamma_0\,(T+\Theta)$ is the damping constant, $\xi_{\kern-.5pt T}=\xi_0\,(T+\Theta)^{-1/2}$ is the magnetic correlation length, and $\Theta$ is the Curie-Weiss temperature. We obtain the best fit to all the normal-state data (Figs.\,1b, 2, and 3) for $\chi_0\kern-.7pt=\kern-.5pt(3.8\kern-.5pt\pm\kern-.5pt1.0)\cdot\kern-.5pt10^4\,\mu_\text{B}^2$\,K/eV, $\Gamma_0=(0.14\pm0.04)$\,meV/K, $\Theta=(30\pm10)$\,K, and $\xi_0=(163\pm20)$\,\AA\,K$^{1/2}$, shown as dashed lines in Fig.\,2a. The deviation of the model from the experimental data at high energies can possibly be explained by the presence of multiple bands in the system, which shifts the maximum of $\chi''_{60\,\text{K}}(\mathbf{Q}_\text{AFM},\,\omega)$ to a higher value of $\sim$\,$20$\,meV. The total spectral weight at 60\,K, integrated over \qq\ and $\omega$ up to 35\,meV is $\chi''_{60\,\textup{K}}=0.17\,\mu_\textup{B}^2$/f.u., and is thus comparable to underdoped \ybco\ \cite{FongBourges00}. The net resonance intensity, on the other hand, amounts to $\chi''_{\textup{res}}=\chi''_{4\,\textup{K}}-\chi''_{60\,\textup{K}}=0.013\,\mu_\textup{B}^2$/f.u., which is 3\,--\,5 times smaller than in \ybco\ \cite{FongBourges00}.

From Fig.\,2 we can define three energy intervals: The spin gap below $\sim$\,3\,meV, the resonance region between $\sim$\,3 and $\sim$\,15\,meV, and the region above $\sim$\,15\,meV with no superconductivity-induced changes. In Fig.\,3a we show the evolution of $\chi''(\qaf,\omega)$ at the representative energies 3, 9.5 and 16\,meV for temperatures up to 280\,K. We observe a smooth increase upon cooling down to \Tc\ at all three energies. While at 16\,meV the intensity also evolves smoothly across \Tc, there are pronounced anomalies at 3 and 9.5\,meV, indicating the abrupt gap opening. We note that there is no indication of a pseudogap opening above \Tc, which is consistent with the linear behavior of $\chi''(\qq,\omega)$ at small $\omega$ (Fig.\,2).

However, since the superconducting gap decreases upon heating to \Tc\ \cite{EvtushinskyInosov09, TerashimaSekiba09}, it does not suffice to study the $T$-dependence of $\chi''(\qq,\omega)$ at a fixed energy. Hence, we investigated the evolution of the resonance peak by performing energy scans at several temperatures below \Tc\ (Fig.\,3b). An important result is that \Eres\ decreases upon heating as well, and it follows the same functional dependence as \deltaSC\ with remarkable precision, that is $\Eres(T)\propto\deltaSC(T)$ (Fig.\,3c).

A comprehensive summary of our data in the $\omega$-$T$ plane is shown in Fig.\,3d. An extended animation thereof, including the \qq-dependence, is presented in the Supplementary Information. As indicated by the vertical bar, the resonance maximum always remains inside the $2\deltaSC$ gap, while its tail might extend beyond.

What are the implications of our results for the physics and in particular the superconducting mechanism of the iron arsenides? We begin by comparing the normal-state spin excitations of BaFe$_{1.85}$Co$_{0.15}$As$_2$ to those of the cuprates. Remarkably, the overall magnitude of $\chi''(\qq,\omega)$ is similar in both families \cite{FongBourges00, LipscombeVignolle09}.  However, the cuprate spectra exhibit anomalous features such as a ``spin pseudogap'' \cite{RossatMignod91, LeeYamada03} and a broad peak reminiscent of the resonant mode in the normal state \cite{FongBourges00}. In contrast, we have shown that the normal-state spin excitation spectrum of BaFe$_{1.85}$Co$_{0.15}$As$_2$ is gapless and can be well described by a simple formula for nearly antiferromagnetic metals described by Moriya in Ref.\,\onlinecite{Moriya85}. This encourages us to use the simple approach towards a spin-excitation mediated pairing in such systems described in Ref. \onlinecite{MonthouxBalatsky91} for a rough comparison of the spin-fermion coupling in arsenides and cuprates. Neglecting material-specific complications like the multi-band character of the iron arsenides, we obtain
\begin{equation}
\lambda\approx-\Biggl[\ln\kern.5pt\frac{k_\text{B}T_\text{c}}{\Gamma_{\kern-.5pt T_\text{c}}(\xi_{\kern-.5pt T_\text{c}}\big/a)^2}\Biggr]^{-1}\!\approx0.2,
\label{lambda}
\end{equation}
for the dimensionless effective coupling constant $\lambda$ between spin excitations and quasiparticles. Although more elaborate calculations have been reported in the meantime \cite{ChubukovPines03}, this rough estimate suggests that the coupling is weaker than in cuprates, where Eq.\,\eqref{lambda} yields $\lambda= 0.35\,-\,0.45$. This finding is consistent with the fact that despite the comparable normal state magnitude of $\chi''(\qq,\omega)$ in iron arsenides and cuprates, \Tc\ and the resonance enhancement of $\chi''(\qq,\omega)$ below \Tc\ are significantly lower in the former.

 In the light of our results, Fermi-liquid based theories like the random phase approximation (RPA) \cite{KorshunovEremin08, MaierGraser09} and related approaches appear much better justified in arsenides than in cuprates. Turning now to the superconducting state, we first note that the impact of superconductivity on the spin excitations can be fully accounted for by the opening of \deltaSC\ and the appearance of the resonance, without qualitative changes to the excitation geometry. Considering the resonance as a bound state within the superconducting gap, $\Eres<2\deltaSC$ is required, and our value of $\Eres=(1.6\pm0.3)\deltaSC$ is in good agreement with the predictions for a sign-reversed $s_{\pm}$-wave gap \cite{KorshunovEremin08, MaierGraser09}. Furthermore, we have shown that \Eres\ follows the same trend as $\deltaSC(T)$ when the gap closes upon heating, as expected from conventional Fermi liquid based approaches. Once more, the simplicity of this behavior is in notable contrast to its counterpart in the cuprates \cite{FongKeimer96}, where the temperature insensitivity of \Eres\ has inspired theories that attribute the spin resonance to a particle-particle bound state \cite{DemlerZhang95} or a collective mode characteristic of a state competing with superconductivity \cite{MorrPines98}.

Finally we mention that for the moment the observed pinning of spin excitations to \qaf\ cannot be reconciled with predictions of incommensurate excitations \cite{MaierGraser09} based on the notion that the nesting vector should deviate from \qaf\ due to electron-doping related Fermi surface changes.

In conclusion, the comprehensive set of data on the spin dynamics in BaFe$_{1.85}$Co$_{0.15}$As$_2$ in the normal and superconducting states we have presented will enable a rigorous assessment of spin-fluctuation-mediated pairing models for the iron arsenides. In particular, based on our absolute-unit calibration of $\chi''(\qq,\omega)$ it will become possible to compare the total exchange energy of the electron system below \Tc\ to the condensation energy determined by specific-heat measurements \cite{DemlerZhang98}. Independent information about the spin-fermion coupling strength can also be derived from a comparison of the measured spin fluctuation spectrum and the fermionic self-energy extracted from photoemission spectroscopy. Although these complementary approaches have yielded important insights into the mechanism of superconductivity in cuprates \cite{DahmHinkov09, WooDai06}, a controlled, commonly accepted theory is still missing. Our data provide tantalizing indications that such a theory may be within reach for the iron arsenides.

\small\noindent\smallskip

\noindent\textbf{Methods} Our sample is a single crystal of BaFe$_{1.85}$Co$_{0.15}$As$_2$ with a mass of $1.0$~g. It was grown with the self-flux method to prevent contaminations, and using a nucleation center. Further growth details are described elsewhere \cite{SunLiu09}. The high crystalline quality was assessed by neutron- and X-ray diffraction measurements. The superconducting transition temperature was determined by SQUID magnetometry to be $\Tc=25$\,K, which corresponds to an optimal doping level according to the phase diagram \cite{ChuAnalytis09}.

We use tetragonal notation and quote the transferred wavevector \qq\ in units of the reciprocal lattice vectors \astar, \bstar\ and \cstar. In this notation, the antiferromagnetic wavevector is $\qaf=\bigl(\frac{1}{2}\,\frac{1}{2}\,1\bigr)$.

The data were collected using the cold triple-axis \textit{Panda} and thermal triple-axis \textit{Puma} spectrometers (FRM-II, Garching, Germany), as well as the 2T spectrometer (LLB, Saclay, France).  The sample was mounted into a standard cryostat with the (110) and (001) directions in the scattering plane. In all cases, pyrolytic graphite monochromators and analyzers were used. Measurements were performed in constant-$k_\textup{f}$ mode, with $k_\textup{f}=1.55$\,\AA$^{-1}$ in conjunction with a Beryllium filter at small $\omega$ and $k_\textup{f}=2.66$\,\AA$^{-1}$ or $k_\textup{f}=4.1$\AA$^{-1}$ with a pyrolytic graphite filter at large $\omega$. Wherever applicable, the background was subtracted from the data, and corrections for the magnetic structure factor and for the energy-dependent fraction of higher-order neutrons were applied. The imaginary part of the dynamical spin susceptibility $\chi''(\qq,\omega)$ was obtained from the scattering function $S(\qq,\omega)$ by the fluctuation-dissipation relation $\chi''(\qq,\omega)=(1-e^{-\hslash\omega/k_\textup{B}T})\,S(\qq,\omega)$.  The datasets measured at different spectrometers or with different experimental settings were scaled by using overlapping energy regions as a reference. The error bars in all figures correspond to one standard deviation of the count rate and do not include the normalization errors.

We put our data on absolute scale by comparing the magnetic scattering intensity to the intensity of acoustic phonons as well as nuclear Bragg peaks after taking care of resolution corrections. This approach is extensively discussed in Ref.\,\onlinecite{FongBourges00} and references therein, from which we also adopt the definition of $\chi''$ as $\operatorname{Tr}\chi''_{\alpha\beta}/3$, where $\chi''_{\alpha\beta}$ is the imaginary part of the generalized susceptibility tensor.

\smallskip
\noindent\textbf{Acknowledgements} The project was supported, in part, by the DFG in the consortium FOR538. We are
grateful to L. Boeri, T.~Dahm, O.~Dolgov, D.~Efremov, I.~Eremin, D.~Evtushinsky, G.~Jackeli, G.~Khaliullin, A.~Yaresko, and R.~Zeyher for stimulating discussions and numerous helpful suggestions.
\smallskip

\noindent\textbf{Author Information} Correspondence and requests for materials should be addressed to V.~H. (\href{mailto:v.hinkov@fkf.mpg.de}{v.hinkov@fkf.mpg.de}).

\bibliographystyle{naturemag}
\footnotesize
\bibliography{BFCA-INS}

\begin{thebibliography}{10}
\expandafter\ifx\csname url\endcsname\relax
  \def\url#1{\texttt{#1}}\fi
\expandafter\ifx\csname urlprefix\endcsname\relax\def\urlprefix{URL }\fi
\providecommand{\bibinfo}[2]{#2}
\providecommand{\eprint}[2][]{\url{#2}}

\bibitem{ChuAnalytis09}
\bibinfo{author}{Chu, J.-H.}, \bibinfo{author}{Analytis, J.~G.},
  \bibinfo{author}{Kucharczyk, C.} \& \bibinfo{author}{Fisher, I.~R.}
\newblock \bibinfo{title}{Determination of the phase diagram of the
  electron-doped superconductor {Ba(Fe$_{1-x}$Co$_x$)$_2$As$_2$}}.
\newblock \emph{\bibinfo{journal}{Phys. Rev.~B}} \textbf{\bibinfo{volume}{79}},
  \bibinfo{pages}{014506} (\bibinfo{year}{2009}).

\bibitem{DrewNiedermayer09}
\bibinfo{author}{Drew, A.~J.} \emph{et~al.}
\newblock \bibinfo{title}{Coexistence of static magnetism and superconductivity
  in {SmFeAsO$_{1-x}$F$_x$} as revealed by muon spin rotation}.
\newblock \emph{\bibinfo{journal}{Nature Mater.}} \textbf{\bibinfo{volume}{8}},
  \bibinfo{pages}{310} (\bibinfo{year}{2009}).

\bibitem{BoeriDolgov08}
\bibinfo{author}{Boeri, L.}, \bibinfo{author}{Dolgov, O.~V.} \&
  \bibinfo{author}{Golubov, A.~A.}
\newblock \bibinfo{title}{Is {LaFeAsO$_{1-x}$F$_x$} an electron-phonon
  superconductor?}
\newblock \emph{\bibinfo{journal}{Phys. Rev. Lett.}}
  \textbf{\bibinfo{volume}{101}}, \bibinfo{pages}{026403}
  (\bibinfo{year}{2008}).

\bibitem{ChristiansonGoremychkin08}
\bibinfo{author}{Christianson, A.~D.} \emph{et~al.}
\newblock \bibinfo{title}{Unconventional superconductivity in
  {Ba$_{0.6}$K$_{0.4}$Fe$_{2}$As$_{2}$} from inelastic neutron scattering}.
\newblock \emph{\bibinfo{journal}{Nature}} \textbf{\bibinfo{volume}{456}},
  \bibinfo{pages}{930} (\bibinfo{year}{2008}).

\bibitem{ChiSchneidewind09}
\bibinfo{author}{Chi, S.} \emph{et~al.}
\newblock \bibinfo{title}{Inelastic neutron-scattering measurements of a
  three-dimensional spin resonance in the {FeAs}-based
  {BaFe$_{1.9}$Ni$_{0.1}$As$_2$} superconductor}.
\newblock \emph{\bibinfo{journal}{Phys. Rev. Lett.}}
  \textbf{\bibinfo{volume}{102}}, \bibinfo{pages}{107006}
  (\bibinfo{year}{2009}).

\bibitem{LumsdenChristianson09}
\bibinfo{author}{Lumsden, M.~D.} \emph{et~al.}
\newblock \bibinfo{title}{Two-dimensional resonant magnetic excitation in
  {BaFe$_{1.84}$Co$_{0.16}$As$_2$}}.
\newblock \emph{\bibinfo{journal}{Phys. Rev. Lett.}}
  \textbf{\bibinfo{volume}{102}}, \bibinfo{pages}{107005}
  (\bibinfo{year}{2009}).

\bibitem{LiChen09}
\bibinfo{author}{Li, S.} \emph{et~al.}
\newblock \bibinfo{title}{Spin gap and magnetic resonance in superconducting
  {BaFe$_{1.9}$Ni$_{0.1}$As$_{2}$}}.
\newblock \emph{\bibinfo{journal}{Phys. Rev.~B}} \textbf{\bibinfo{volume}{79}},
  \bibinfo{pages}{174527} (\bibinfo{year}{2009}).

\bibitem{Parks69}
\bibinfo{editor}{Parks, R.~D.} (ed.) \emph{\bibinfo{title}{Superconductivity}},
  vol.~\bibinfo{volume}{1} (\bibinfo{publisher}{Dekker}, \bibinfo{year}{1969}).

\bibitem{DahmHinkov09}
\bibinfo{author}{Dahm, T.} \emph{et~al.}
\newblock \bibinfo{title}{Strength of the spin-fluctuation-mediated pairing
  interaction in a high-temperature superconductor}.
\newblock \emph{\bibinfo{journal}{Nature Phys.}} \textbf{\bibinfo{volume}{5}},
  \bibinfo{pages}{217} (\bibinfo{year}{2009}).

\bibitem{FongBourges00}
\bibinfo{author}{Fong, H.~F.} \emph{et~al.}
\newblock \bibinfo{title}{Spin susceptibility in underdoped
  {YBa$_2$Cu$_3$O$_{6+x}$}}.
\newblock \emph{\bibinfo{journal}{Phys. Rev.~B}} \textbf{\bibinfo{volume}{61}},
  \bibinfo{pages}{14773} (\bibinfo{year}{2000}).

\bibitem{LipscombeVignolle09}
\bibinfo{author}{Lipscombe, O.~J.}, \bibinfo{author}{Vignolle, B.},
  \bibinfo{author}{Perring, T.~G.}, \bibinfo{author}{Frost, C.~D.} \&
  \bibinfo{author}{Hayden, S.~M.}
\newblock \bibinfo{title}{Emergence of coherent magnetic excitations in the
  high temperature underdoped {La$_{2-x}$Sr$_x$CuO$_4$} superconductor at low
  temperatures}.
\newblock \emph{\bibinfo{journal}{Phys. Rev. Lett.}}
  \textbf{\bibinfo{volume}{102}}, \bibinfo{pages}{167002}
  (\bibinfo{year}{2009}).

\bibitem{Moriya85}
\bibinfo{author}{Moriya, T.}
\newblock \emph{\bibinfo{title}{Spin Fluctuations in Itinerant Electron
  Magnetism}} (\bibinfo{publisher}{Springer-Verlag}, \bibinfo{address}{Berlin
  Heidelberg}, \bibinfo{year}{1985}).

\bibitem{NormanPines05}
\bibinfo{author}{Norman, M.~R.}, \bibinfo{author}{Pines, D.} \&
  \bibinfo{author}{Kallin, C.}
\newblock \bibinfo{title}{The pseudogap: friend of foe of high {$T_c$}?}
\newblock \emph{\bibinfo{journal}{Adv. Phys.}} \textbf{\bibinfo{volume}{54}},
  \bibinfo{pages}{715--733} (\bibinfo{year}{2005}).

\bibitem{RossatMignod91}
\bibinfo{author}{Rossat-Mignod, J.} \emph{et~al.}
\newblock \bibinfo{title}{Neutron scattering study of the
  {YBa$_2$Cu$_3$O$_{6+x}$} system}.
\newblock \emph{\bibinfo{journal}{Physica C: Superconductivity}}
  \textbf{\bibinfo{volume}{185}}, \bibinfo{pages}{86} (\bibinfo{year}{1991}).

\bibitem{LeeYamada03}
\bibinfo{author}{Lee, C.~H.}, \bibinfo{author}{Yamada, K.},
  \bibinfo{author}{Hiraka, H.}, \bibinfo{author}{{Venkateswara Rao}, C.~R.} \&
  \bibinfo{author}{Endoh, Y.}
\newblock \bibinfo{title}{Spin pseudogap in {La$_{2-x}$Sr$_x$CuO$_4$} studied
  by neutron scattering}.
\newblock \emph{\bibinfo{journal}{Phys. Rev.~B}} \textbf{\bibinfo{volume}{67}},
  \bibinfo{pages}{134521} (\bibinfo{year}{2003}).

\bibitem{TranquadaSternlieb95}
\bibinfo{author}{Tranquada, J.~M.}, \bibinfo{author}{Sternlieb, B.~J.},
  \bibinfo{author}{Axe, J.~D.}, \bibinfo{author}{Nakamura, Y.} \&
  \bibinfo{author}{Uchida, S.}
\newblock \bibinfo{title}{Evidence for stripe correlations of spins and holes
  in copper oxide superconductors}.
\newblock \emph{\bibinfo{journal}{Nature}} \textbf{\bibinfo{volume}{375}},
  \bibinfo{pages}{561} (\bibinfo{year}{1995}).

\bibitem{KorshunovEremin08}
\bibinfo{author}{Korshunov, M.~M.} \& \bibinfo{author}{Eremin, I.}
\newblock \bibinfo{title}{Theory of magnetic excitations in iron-based layered
  superconductors}.
\newblock \emph{\bibinfo{journal}{Phys. Rev.~B}} \textbf{\bibinfo{volume}{78}},
  \bibinfo{pages}{140509} (\bibinfo{year}{2008}).

\bibitem{MaierGraser09}
\bibinfo{author}{Maier, T.~A.}, \bibinfo{author}{Graser, S.},
  \bibinfo{author}{Scalapino, D.~J.} \& \bibinfo{author}{Hirschfeld, P.}
\newblock \bibinfo{title}{Neutron scattering resonance and the iron-pnictide
  superconducting gap}.
\newblock \emph{\bibinfo{journal}{Phys. Rev.~B}} \textbf{\bibinfo{volume}{79}},
  \bibinfo{pages}{134520} (\bibinfo{year}{2009}).

\bibitem{SatoAso01}
\bibinfo{author}{Sato, N.~K.} \emph{et~al.}
\newblock \bibinfo{title}{Strong coupling between local moments and
  superconducting `heavy' electrons in {UPd$_2$Al$_3$}}.
\newblock \emph{\bibinfo{journal}{Nature}} \textbf{\bibinfo{volume}{410}},
  \bibinfo{pages}{340} (\bibinfo{year}{2001}).

\bibitem{Eschrig06}
\bibinfo{author}{Eschrig, M.}
\newblock \bibinfo{title}{The effect of collective spin-1 excitations on
  electronic spectra in high-{$T_\text{c}$} superconductors}.
\newblock \emph{\bibinfo{journal}{Adv. Phys.}} \textbf{\bibinfo{volume}{55}},
  \bibinfo{pages}{47} (\bibinfo{year}{2006}).

\bibitem{Anderson07}
\bibinfo{author}{Anderson, P.~W.}
\newblock \bibinfo{title}{Is there a glue in cuprate superconductors}.
\newblock \emph{\bibinfo{journal}{Science}} \textbf{\bibinfo{volume}{316}},
  \bibinfo{pages}{1705} (\bibinfo{year}{2007}).

\bibitem{KamiharaWatanabe08}
\bibinfo{author}{Kamihara, Y.}, \bibinfo{author}{Watanabe, T.},
  \bibinfo{author}{Hirano, M.} \& \bibinfo{author}{Hosono, H.}
\newblock \bibinfo{title}{Iron-based layered superconductor
  {La[O$_{1-x}$F$_x$]FeAs} ($x = 0.05$--$0.12$) with {$T_\text{c} = 26$\,K}}.
\newblock \emph{\bibinfo{journal}{J. Am. Chem. Soc.}}
  \textbf{\bibinfo{volume}{130}}, \bibinfo{pages}{3296} (\bibinfo{year}{2008}).

\bibitem{SefatJin08}
\bibinfo{author}{Sefat, A.~S.} \emph{et~al.}
\newblock \bibinfo{title}{Superconductivity at {22\,K} in {Co}-doped
  {BaFe$_2$As$_2$} crystals}.
\newblock \emph{\bibinfo{journal}{Phys. Rev. Lett.}}
  \textbf{\bibinfo{volume}{101}}, \bibinfo{pages}{117004}
  (\bibinfo{year}{2008}).

\bibitem{MazinSingh08}
\bibinfo{author}{Mazin, I.~I.}, \bibinfo{author}{Singh, D.~J.},
  \bibinfo{author}{Johannes, M.~D.} \& \bibinfo{author}{Du, M.~H.}
\newblock \bibinfo{title}{Unconventional superconductivity with a sign reversal
  in the order parameter of {LaFeAsO$_{1-x}$F$_x$}}.
\newblock \emph{\bibinfo{journal}{Phys. Rev. Lett.}}
  \textbf{\bibinfo{volume}{101}}, \bibinfo{pages}{057003}
  (\bibinfo{year}{2008}).

\bibitem{KurokiOnari08}
\bibinfo{author}{Kuroki, K.} \emph{et~al.}
\newblock \bibinfo{title}{Unconventional pairing originating from the
  disconnected {Fermi} surfaces of superconducting {LaFeAsO$_{1-x}$F$_x$}}.
\newblock \emph{\bibinfo{journal}{Phys. Rev. Lett.}}
  \textbf{\bibinfo{volume}{101}}, \bibinfo{pages}{087004}
  (\bibinfo{year}{2008}).

\bibitem{FongKeimer96}
\bibinfo{author}{Fong, H.~F.}, \bibinfo{author}{Keimer, B.},
  \bibinfo{author}{Reznik, D.}, \bibinfo{author}{Milius, D.~L.} \&
  \bibinfo{author}{Aksay, I.~A.}
\newblock \bibinfo{title}{Polarized and unpolarized neutron-scattering study of
  the dynamical spin susceptibility of {YBa$_2$Cu$_3$O$_7$}}.
\newblock \emph{\bibinfo{journal}{Phys. Rev.~B}} \textbf{\bibinfo{volume}{54}},
  \bibinfo{pages}{6708} (\bibinfo{year}{1996}).

\bibitem{EvtushinskyInosov09}
\bibinfo{author}{Evtushinsky, D.~V.} \emph{et~al.}
\newblock \bibinfo{title}{Momentum dependence of the superconducting gap in
  {Ba$_{1-x}$K$_x$Fe$_2$As$_2$}}.
\newblock \emph{\bibinfo{journal}{Phys. Rev.~B}} \textbf{\bibinfo{volume}{79}},
  \bibinfo{pages}{054517} (\bibinfo{year}{2009}).

\bibitem{TerashimaSekiba09}
\bibinfo{author}{Terashima, K.} \emph{et~al.}
\newblock \bibinfo{title}{Fermi surface nesting induced strong pairing in
  iron-based superconductors}.
\newblock \emph{\bibinfo{journal}{Proc. Natl. Acad. Sci. USA}}
  \textbf{\bibinfo{volume}{106}}, \bibinfo{pages}{7330} (\bibinfo{year}{2009}).

\bibitem{SamuelyPribulova09}
\bibinfo{author}{Samuely, P.} \emph{et~al.}
\newblock \bibinfo{title}{Point contact {Andreev} reflection spectroscopy of
  superconducting energy gaps in 122-type family of iron pnictides}.
\newblock \emph{\bibinfo{journal}{Physica C: Superconductivity}}
  \textbf{\bibinfo{volume}{469}}, \bibinfo{pages}{507} (\bibinfo{year}{2009}).

\bibitem{YinZech09}
\bibinfo{author}{Yin, Y.} \emph{et~al.}
\newblock \bibinfo{title}{Scanning tunneling spectroscopy and vortex imaging in
  the iron pnictide superconductor {BaFe$_{1.8}$Co$_{0.2}$As$_2$}}.
\newblock \emph{\bibinfo{journal}{Phys. Rev. Lett.}}
  \textbf{\bibinfo{volume}{102}}, \bibinfo{pages}{097002}
  (\bibinfo{year}{2009}).

\bibitem{MonthouxBalatsky91}
\bibinfo{author}{Monthoux, P.}, \bibinfo{author}{Balatsky, A.~V.} \&
  \bibinfo{author}{Pines, D.}
\newblock \bibinfo{title}{Toward a theory of high-temperature superconductivity
  in the antiferromagnetically correlated cuprate oxides}.
\newblock \emph{\bibinfo{journal}{Phys.~Rev.~Lett.}}
  \textbf{\bibinfo{volume}{67}}, \bibinfo{pages}{3448} (\bibinfo{year}{1991}).

\bibitem{ChubukovPines03}
\bibinfo{note}{Chubukov, A. V., Pines, D., Schmalian, J. A spin fluctuation
  model for $d$-wave superconductivity. \emph{The physics of superconductors.}
  Eds. Bennemann, K. H. and Ketterson, J. B. (Springer-Verlag 2003)}.

\bibitem{DemlerZhang95}
\bibinfo{author}{Demler, E.} \& \bibinfo{author}{Zhang, S.-C.}
\newblock \bibinfo{title}{Theory of the resonant neutron scattering of
  {high-$Tc$} superconductors}.
\newblock \emph{\bibinfo{journal}{Phys. Rev. Lett.}}
  \textbf{\bibinfo{volume}{75}}, \bibinfo{pages}{4126--4129}
  (\bibinfo{year}{1995}).

\bibitem{MorrPines98}
\bibinfo{author}{Morr, D.~K.} \& \bibinfo{author}{Pines, D.}
\newblock \bibinfo{title}{The resonance peak in cuprate superconductors}.
\newblock \emph{\bibinfo{journal}{Phys. Rev. Lett.}}
  \textbf{\bibinfo{volume}{81}}, \bibinfo{pages}{1086--1089}
  (\bibinfo{year}{1998}).

\bibitem{DemlerZhang98}
\bibinfo{author}{Demler, E.} \& \bibinfo{author}{Zhang, S.}
\newblock \bibinfo{title}{{Quantitative test of a microscopic mechanism of
  high-temperature superconductivity}}.
\newblock \emph{\bibinfo{journal}{{Nature}}} \textbf{\bibinfo{volume}{{396}}},
  \bibinfo{pages}{{733--735}} (\bibinfo{year}{{1998}}).

\bibitem{WooDai06}
\bibinfo{author}{Woo, H.} \emph{et~al.}
\newblock \bibinfo{title}{{Magnetic energy change available to superconducting
  condensation in optimally doped YBa$_2$Cu$_3$O$_{6.95}$}}.
\newblock \emph{\bibinfo{journal}{{Nature Phys.}}}
  \textbf{\bibinfo{volume}{{2}}}, \bibinfo{pages}{{600--604}}
  (\bibinfo{year}{{2006}}).

\bibitem{SunLiu09}
\bibinfo{note}{Sun, D. L., Liu, Y., Park, J. T. and Lin, C. T., to be published
  in J. Mater. Sci. (2009).}

\end{thebibliography}

\clearpage\normalsize

\begin{widetext}

\begin{flushright}\LARGE\textsf{SUPPLEMENTARY INFORMATION}\end{flushright}\vspace{-1em}
\hrulefill

\renewcommand{\thefigure}{S1}
\begin{figure}[b]
\includegraphics[width=0.8\textwidth]{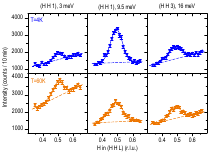}
\caption{Several raw $Q$-scans measured in the superconducting state (top row, $T=4$\,K) and in the normal state (bottom row, $T=60$\,K) at three different energies: 3\,meV, 9.5\,meV, and 16\,meV. The solid lines represent Gaussian fits with a linear background. The background is indicated by dashed lines.\vspace{-0.5em}}
\label{Q-scans}
\end{figure}

\section{Data processing}

\noindent The presented data were collected in the fixed $k_\text{f}$ mode ($k_\text{f}$ being the final momentum of the scattered neutrons) using three different triple-axis spectrometers. The low-energy part of the spectrum was measured with $k_\text{f}=1.549$\,\AA$^{-1}$ at the cold neutron spectrometer Panda at the FRM-II research reactor in Garching, Germany, whereas the higher-energy data were collected with $k_\text{f}=2.662,~3.840\text{, and }4.098$\,\AA$^{-1}$ at the thermal neutron spectrometers Puma (FRM II) and 2T (LLB, Saclay, France). Measurements at different spectrometers are indicated by the different shapes of the data points in the figures. (002) pyrolytic graphite monochromators and analyzers were used in all the experiments. In the cold neutron experiment, a Be filter was used on $k_\text{f}$ to eliminate the contamination from the higher-order neutrons. In the thermal neutron experiment, a pyrolytic graphite filter was used for the same purpose.

In Fig.\,S1, several representative $Q$-scans across the AFM wavevector are shown. One can see that the signal is well fitted by a single Gaussian peak with a linear background, showing no signatures of incommensurability within the energy range of up to $2\Delta$. The measured signal has been corrected to account for the energy-dependent fraction of higher-order neutrons. The intensity at $\mathbf{Q}_\text{AF}$ shown in Fig.\,1a in the paper was obtained from such $Q$-scans by subtracting the fitted background line from the raw data. At those energies, where no full $Q$-scans were available, the background was estimated as the average intensity measured in two points on both sides of the peak. Next, the data measured at different values of the out-of-plane component of the wavevector $L$ were scaled by the magnetic structure factor, determined from the direct comparison of resonance intensities at those $L$ values. To put the intensity obtained in different experiments on the same scale, the datasets were scaled by using the overlapping energy regions as a reference. The error bars in all figures correspond to one standard deviation
of the count rate and do not include the uncertainties of these scaling factors.

\section{Temperature, energy, and momentum dependence of the dynamic spin susceptibility}

\noindent By combining the energy and temperature dependence of $\chi''(\mathbf{Q}_\text{AFM},\,\omega)$ presented in Fig.\,2 with the momentum-dependence as given by Eq.\,(1) in the paper, it was possible to reconstruct phenomenologically the full functional dependence of the dynamic spin susceptibility on temperature, momentum, and energy, which we present as an animation in Fig.\,S2 below. The momentum-dependence was checked to be consistent with the measured $\mathbf{Q}$-scans in the whole temperature range, which indicate a broadening with increasing energy in agreement with Eq.\,(1).

\renewcommand{\thefigure}{S2}
\begin{figure}[h]
\animategraphics[controls, loop, width=0.9\textwidth]{2}{resonance_}{1}{17}
\caption{Animation of $\chi''(q,\,\omega)$ as a function of the wavevector $q=|\mathbf{Q}-\mathbf{Q}_\text{AFM}|$, energy $\omega$, and temperature $T$.}
\label{Animation}
\end{figure}

\clearpage

\end{widetext}

\end{document}